\documentclass[journal=ancac3,manuscript=article]{achemso}

\usepackage[version=3]{mhchem}

\usepackage{amsmath}
\usepackage{graphicx}

\title[\texttt{achemso} demonstration]
{Bilayer Graphene Lateral Contacts for DNA Sequencing}

\author{Yuhui He}
\email{he.yuhui.ime@gmail.com, he.yuhui@sanken.osaka-u.ac.jp}
\affiliation[ISIR, Osaka University]
{The Institute of Scientific and Industrial Research, Osaka University, 8-1 Mihogaoka, Ibaraki, Osaka 567-0047, Japan}
\author{Makusu Tsutsui}
\email{makusu32@sanken.osaka-u.ac.jp}
\affiliation[ISIR, Osaka University]
{The Institute of Scientific and Industrial Research, Osaka University, 8-1 Mihogaoka, Ibaraki, Osaka 567-0047, Japan}
\author{Ralph H. Scheicher}
\email{ralph.scheicher@fysik.uu.se}
\affiliation[Uppsala University]
{Division of Materials Theory, Department of Physics and Astronomy, Box 516, Uppsala University, SE-751 20 Uppsala, Sweden}
\author{Masateru Taniguchi}
\email{taniguti@sanken.osaka-u.ac.jp}
\affiliation[ISIR, Osaka University]
{The Institute of Scientific and Industrial Research, Osaka University, 8-1 Mihogaoka, Ibaraki, Osaka 567-0047, Japan}
\author{Tomoji Kawai}
\email{kawai@sanken.osaka-u.ac.jp}
\affiliation[ISIR, Osaka University]
{The Institute of Scientific and Industrial Research, Osaka University, 8-1 Mihogaoka, Ibaraki, Osaka 567-0047, Japan}

\begin{document}
\begin{abstract}
Translocation of DNA through a nanopore with embedded electrodes is at the centre of new rapid inexpensive sequencing methods which allow distinguishing the four nucleobases by their different electronic structure. However, the subnanometer separation between nucleotides in DNA requires  ultra-sharp probes. Here, we propose a device architecture consisting of a nanopore formed in bilayer graphene, with the two layers acting as separate electrical contacts. The 0.34 nm interlayer distance of graphene is ideally suited for electrical
coupling to a single nucleobase, avoiding the difficulty of fabricating probes with subnanometer precision. The top and bottom graphene electrodes contact the target molecule from the same lateral side, removing the orders-of-magnitude tunneling current variations between smaller pyrimidine bases and larger purine bases. We
demonstrate that incorporating techniques for molecular manipulation enables the proposed device to sequence single-stranded DNA and that it offers even the prospect of sequencing double-stranded DNA.
\end{abstract}

\begin{center}
\end{center}

Electrical monitoring of DNA translocation through a nanopore has been proposed as a fast, low-cost and high-throughput sequencing technique \cite{Kasianowicz1996,Deamer1999,Meller2000,Meller2001}, and the past decade has witnessed tremendous progress towards that goal \cite{Bayley2005, Zwolak2008, Branton2008, Zwolak2005, Lagerqvist2006, Tsutsui2010, Huang2010, Ivanov2011, Venkatesan2011, Lieber2012}. Recently, it has been suggested \cite{Postma2010} that one could create a nanogap in graphene and use the resulting edges as transverse electrodes for DNA sequencing. This proposal holds promise for detecting each individual nucleobase on a DNA strand, since graphene edges can be considered to exhibit the ultimate sharpness which a transverse electrical probe could possess. Crucial experimental progress was independently established by several research groups who successfully demonstrated DNA translocation through fabricated graphene nanopores \cite{Schneider2010,Merchant2010,Garaj2010,Venkatesan2012}.

Although graphene nanogaps/nanopores may turn out to be the most feasible and efficient way for sequencing, several major challenges still remain to be addressed. One of the biggest difficulties is to define and align electrical contacts on a nanopore-embedded graphene sheet. If we resort to a conventional 2-dimensional (2-D) in-plane electrode structure, the gap between the probes would need to be very small, \textit{i.e.}, about 2 nm, to ensure that the nucleotides under interrogation can bridge the two electrical contacts. Otherwise, the tunneling current would drop below any measurable range, since it decreases exponentially with the distance between electrodes and molecule. However, such a precise shaping of in-plane graphene electrodes separated by a nanopore is expected to be technically extremely challenging.

\begin{figure}
   \begin{center}
      \includegraphics*[width=12cm]{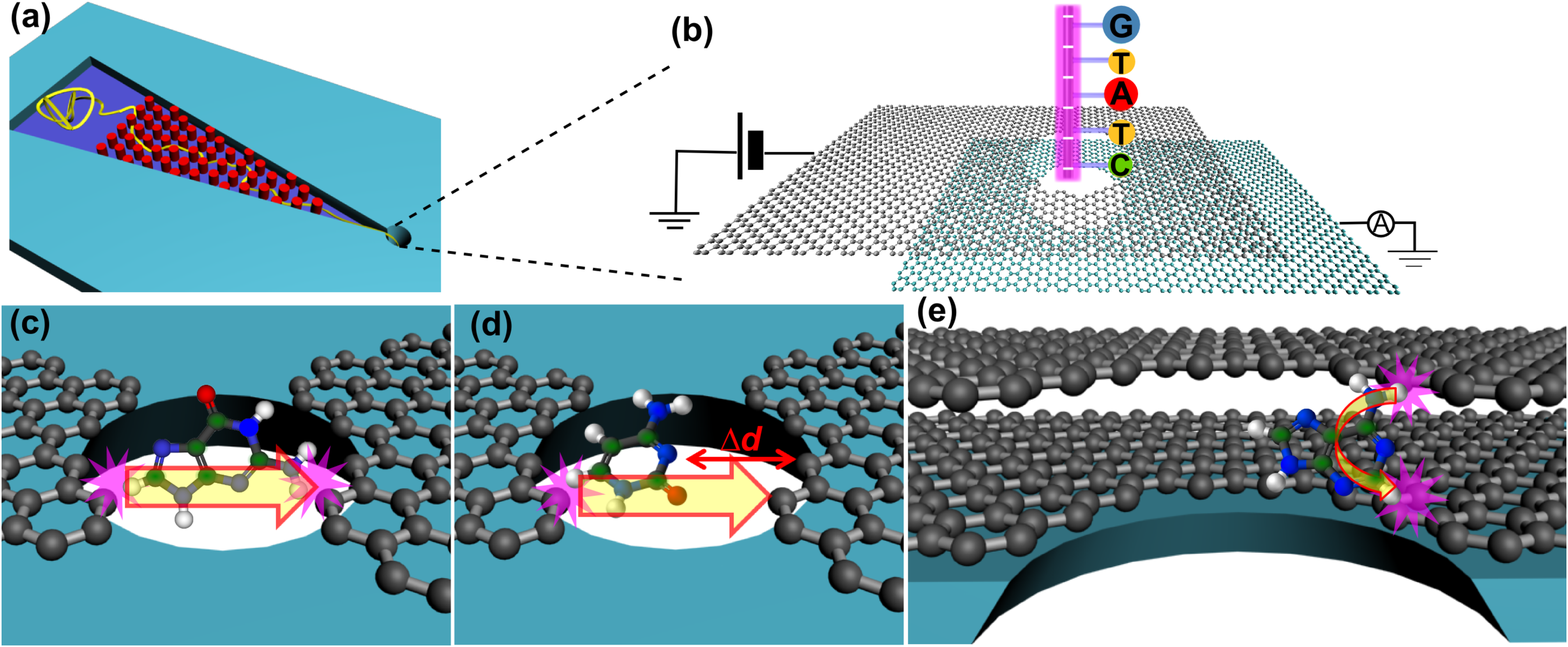}
      \caption{(a) A schematic of our proposed nanopore sequencing system where an array of nanopillars is fabricated for manipulating DNA before entering the nanopore. (b) Magnified view of embedded bilayer graphene electrodes for DNA sequencing, with top and bottom layers as two separate electrical contacts.
      Before DNA entering the nanopore, the nanopillar array serves for stretching the polymer so that the molecule would thread through in one-fold manner; after captured into the pore, the blockage of the transmembrane ionic current triggers a switch of nanopiller surface properties and consequently a much stronger dragging force is exerted on the DNA where overstretch is expected to be induced; then, the tunneling conductance measured by the bilayer graphene electrodes serves as the electrical signature of nucleotides on the overstretched DNA. (c) and (d) Electrical interrogation of guanine and cytosine bases using in-plane graphene electrode structure. Here backbone atoms have been omitted to give a clearer comparison between the two measurements. The coupling of the molecule with the graphene contacts has been marked by the pink lightening symbols. The directions of tunneling currents are schematically characterized by yellow arrows. A nanopore fitting the size of a guanine would be $\Delta d$ larger for that of a cytosine, as characterized in (d). (e) Electrical interrogation of adenine using bilayer graphene contacts where the coupling and electrical current direction are marked in similar way.}
      \label{fig1}
   \end{center}
\end{figure}

Here, we propose a novel architecture to address and potentially solve this issue, namely by using bilayer graphene as the electrical reading elements inside a nanopore as illustrated in \ref{fig1}(a) and (b): an array of nanopillars is mounted with the purpose to stretch the DNA molecule before it enters the nanopore, while bilayer graphene is embedded inside the nanopore as electrical contacts. The uniqueness of the proposed device is that here the top and bottom graphene layers serve as two separate electrical contacts, and the two contacts probe the target molecule from the same lateral sides. Several advantages are expected to be drawn from adopting this bilayer graphene lateral contacts (in the following abbreviated as BGLC) design. First, the interlayer distance between the two van-der-Waals-bound graphene sheets amounts naturally to 0.34 nm, which is less than (for single-stranded DNA) or equal (for double-stranded DNA) to the spacing between adjacent nucleotides, ensuring that no more than one nucleotide should be in contact with both electrodes at a given time, thus readily achieving single-base resolution. This self-assembly of the bilayer graphene to lateral contacts separated by 0.34 nm may be regarded as an important benefit compared to previous proposals which all require the fabrication of graphene-gaps or -ribbons with subnanometer precision \cite{Postma2010,Prezhdo2010,Min2011,Saha2011}. These enormous difficulties in fabrication have so far prevented the experimental realization of electrically detecting individual nucleobases through graphene nanogaps or nanoribbons. We emphasize that those formidable demands on processing precision are not required for our proposed design which should thus allow a much more straightforward implementation. Second, the substantial uncertainty of measurements caused by the variation in size of the different nucleobases in the conventional 2-D transverse electrode structure is also circumvented in our design. In the original proposal, the tunneling conductance through individual nucleotides was thought to be determined by the densities of states (DOS) of the nucleobases and the associated distinction could be utilized as electrical signatures of different nucleotides \cite{Zwolak2005}. But, as shown in \ref{fig1}(c) and (d), the inherent volume difference of the four nucleobase types would lead to significant fluctuations of the coupling strength between target molecule and electrical contacts, resulting in orders-of-magnitude variations in the measured tunneling conductance in the low-bias region. This can be easily understood from the following formula which estimates the tunneling conductance through a molecule (here, a nucleobase) within the linear-response regime \cite{Krems2009,Tsutsui2011}:
\begin{equation}
\label{2D_conductance}
G_0 = \frac{2e^2}{h} \frac{\Gamma_L \Gamma_R}{(\varepsilon_f - \varepsilon_0)^2}
\end{equation}
where $\Gamma_{L(R)}$ represents the coupling between the nucleobase and the left (right) probe, $\varepsilon_f$ is the Fermi energy of graphene, and $\varepsilon_0$ is the energy of the highest occupied molecular orbital (HOMO) or lowest unoccupied molecular orbital (LUMO) of the nucleobase. If $\Gamma_{L/(R)}$ could be assumed to remain constant for different nucleotides, the different HOMO (LUMO) locations of adenine, guanine, cytosine, and thymine (ADE, GUA, CYT, and THY) would indeed determine the magnitude of their respective conductance. However, $\Gamma_{L/(R)}$ is extremely sensitive to the distance between molecule and electrodes. The size variation of different nucleobases unavoidably causes changes in the molecule--electrodes distance, thus causing large fluctuations in $\Gamma_{L/(R)}$ and consequently in the measured tunneling conductance \cite{Zwolak2008}. Contrary to this, in our design, the top and bottom graphene electrodes contact the to-be-scanned nucleotide from the \textit{same} lateral side, as illustrated in \ref{fig1}(e), rather than from two opposite sides, as shown in \ref{fig1}(c) and (d). Therefore, the size variation of the nucleobase under interrogation is not affecting the coupling and the tunneling conductance is now solely governed by the DOS of the nucleobase. We are going to further demonstrate that by incorporating a certain type of molecule manipulation technology, optimized nucleotide discrimination would be attained for over-stretched DNA threading through the nanopore.

Last but not least, the above analysis of lateral contacts implies that double-stranded DNA (dsDNA) could be sequenced directly by a careful design using our BGLC. By incorporating a suitable DNA manipulation technology, only one strand could be kept sufficiently close to BGLC while the other strand would always remain further away. Since the tunneling current depends exponentially on the distance of individual base from the electrodes, the strand close to BGLC will by far dominate the electrical output signal and thus get sequenced. This would be an overwhelming advantage compared to current sequencing proposals based on single strands which are severely restrained by the self-hybridization effect \cite{Kowalczyk2010}. We are going to further explore this prospect and discuss the required protocol below.

Despite the advantages mentioned above, one crucial demand by our bilayer graphene contacts is that the interlayer conductance of graphene $g_0$ should be as small as possible so that the effective signal, which is the tunneling current caused by passing nucleotides, would not be immersed in the fluctuation of a huge $g_0$. This requirement may be fulfilled by fabricating the bilayer with a relative twist with which the two layers are predicted to be electrically isolated \cite{Mele2010,Bistritzer2010}. Experimentally, adjacent layers that were misoriented by only a few degrees when grown epitaxially on SiC were found to be very weakly electrically coupled \cite{Sprinkle2009,Luican2011}. Thus the intrinsic interlayer conductance between adjacent graphene sheets might not be a severe obstacle for our proposal.

\begin{figure}
   \begin{center}
      \includegraphics*[width=6cm]{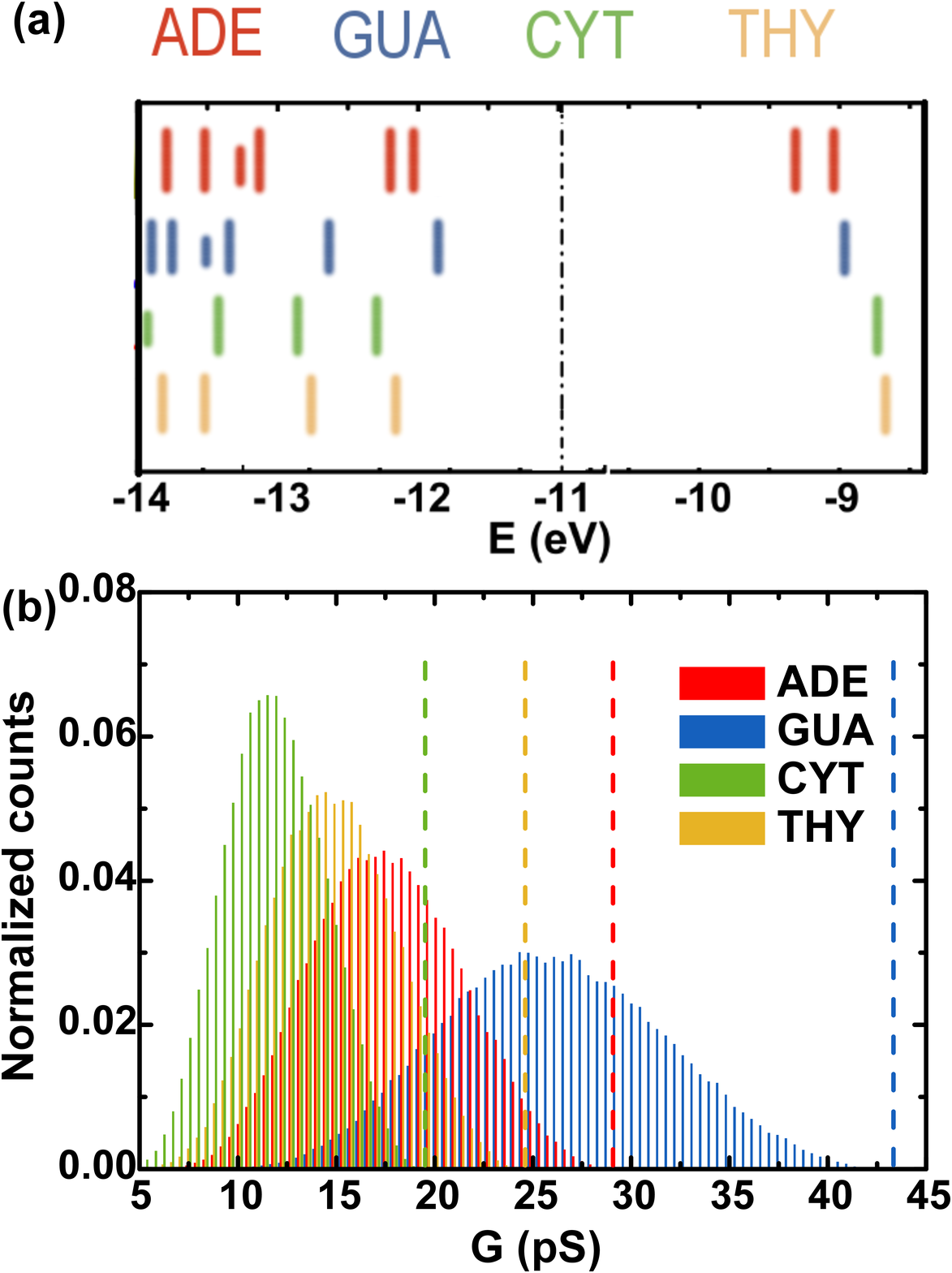}
      \caption{(a) Molecular orbital energies of the four types of nucleotides in the range of interest. The Fermi level of graphene is indicated by a dashed-dotted line. (b) Normalized distribution of tunneling conductance through individual nucleotides from the bilayer graphene electrodes. Here the full-width at half maximum of the conductance distribution is determined by the standard deviation of nucleotide location in the aqueous environment. The dashed lines denote maximum conductance $G_0$ of the four nucleotides when they dwell at the optimized position between the top and bottom graphene layer contacts.
      }
      \label{fig2}
   \end{center}
\end{figure}

\ref{fig2}(a) plots the molecular orbitals of the four types of nucleotides with the Fermi energy of graphene indicated by a dashed-dotted line in the graph. It can be seen that HOMO rather than LUMO would dominate the conductance when using graphene electrodes. The above results were obtained by using extended H\"{u}ckel model and YAEHMOP package, and similar findings were shown in \textit{ab initio} calculations \cite{Prasongkit2011,Saha2011}. Moreover, the locations of the HOMO of different nucleotides shown in this figure would determine the relative order of maximum tunneling conductance of the four bases.

For an estimation of the tunneling conductance through the translocating polynucleotide, we adopt the single energy level model for individual nucleotides \cite{Krems2009} shown in \ref{2D_conductance}, where $\Gamma_{L(R)}$ has been replaced by $\Gamma_{T(B)}$, which is the coupling between the nucleotide and the top (bottom) layer of graphene probe, and $\varepsilon_0$ is now the energy of the HOMO of a given nucleotide. Then, the overall conductance by the whole strand is the sum of contributions from each nucleotide \cite{Postma2010}:
\begin{equation}
\label{tunneling_conductance}
G = \sum_{i=1}^{N}G_0^ie^{-\kappa \left | \vec{r}_i - \vec{r}_0 \right |}
\end{equation}
In the above expression $\kappa = 1.1$ \AA$^{-1}$ is the decay constant of graphene, $G_0^i$ is the maximum tunneling conductance of the $i^{th}$ nucleotide calculated by \ref{2D_conductance}, $\vec{r}_0$ is the corresponding center-of-mass position of that nucleobase when optimally coupled to the electrodes, thus giving rise to the maximum tunneling conductance $G_0^i$, while $\vec{r}_i$ is the actual center-of-mass location of the $i^{th}$ nucleobase during the translocation. In this way, conductance of each base on the target strand is estimated according to the tunneling distance between the base and graphene contacts \cite{Postma2010}.

Let us first discuss the tunneling conductance of single nucleotides dwelling in the bilayer graphene nanopore. The maximum conductances $G_0$ of ADE, GUA, THY and CYT are characterized by dashed lines in \ref{fig2}(b) which due to the distribution of the HOMO energies for four nucleotides are well separated and clearly distinguishable. However, the measured conductance in any real experiment would naturally exhibit significantly expanded distribution curves since there exists stochastic variation in the molecular positions ($\Delta\vec{r}$) during the process of electrical measurement. We estimate $\sqrt{<\Delta\vec{r}^2>} \approx 0.5 $ \AA\ in the confining nanopore based on molecular dynamic (MD) simulations \cite{Phillips2005} (an animation is included in the supporting material), and present the normalized distributions of tunneling conductance with fluctuation effects in \ref{fig2}(b). Although there are overlaps, the distributions for each base remain still discernible. From the experimental point of view, the overlaps indicate that a single measurement would not be sufficient for distinguishing the four nucleobases; rather it calls for several independent measurements and subsequent discriminating based on statistic analysis \cite{Lagerqvist2006,Tsutsui2010}. According to the formalism developed by Lagerqvist \textit{et al.}, about 52 measurements would yield 99\% accuracy (a detailed derivation is provided in supporting materials).

For the whole strand sequencing, we propose that an over-stretching of the DNA molecule within the nanopore is essential to our BGLC system. The physical mechanism is that DNA polymers are highly flexible in the solution. That is, in the absence of any regulation of molecule conformation those adjacent nucleotides on the target strand would experience very large variation of couplings with the lateral graphene probes when passing them. The resulting huge fluctuations in the tunneling conductance would destroy any nucleobase identification effort since the conductance is extremely sensitive to the change of the tunneling barrier. On the other hand, a straightened backbone line of over-stretched DNA could lead to more uniform and improved coupling between each passing-by nucleotide and bilayer probes. This is clearly demonstrated in Fig.S1 within the supporting material.

\begin{figure}
   \begin{center}
      \includegraphics*[width=6cm]{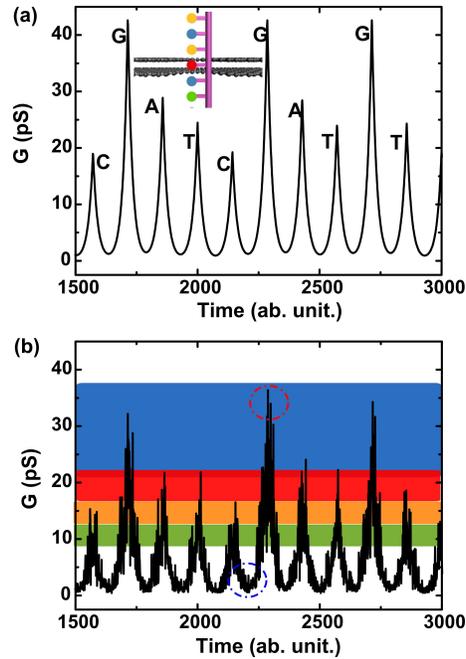}
      \caption{(a) Tunneling conductance versus time, $G(t)$, of a single-stranded DNA translocating through the bilayer graphene nanopore in a highly idealized manner, $i.e.$, the over-stretched polynucleotide is moved with constant speed. The segment shown here contains the base sequence CGATCGATGT. The inset shows schematically the idealized DNA configuration, and lateral bilayer graphene electrodes. (b) $G(t)$ of the same DNA strand taking into account the fluctuations caused by statistical variations in the nucleotide locations during the measurement process. The dashed red circle characterizes a base state, while the blue circle characterizes a gap state.
      }
      \label{fig3}
   \end{center}
\end{figure}

We first discuss a highly idealized case of single-stranded DNA (ssDNA) dynamics and the associated electrical sequencing, where an over-stretched DNA strand slides through the nanopore at constant speed in the absence of any fluctuation effect. This sets the foundation of our proposed approach in the sense of a test whether in the best imaginable scenario the conductance measured by BGLC can be utilized for detecting individual nucleotides on the threaded strand. \ref{fig3}(a) plots the calculated time-dependent tunneling conductance $G(t)$ of segments on a DNA strand with the sequence CGATCGATGT. The inset illustrates the highly idealized configuration of overstretched ssDNA in the bilayer graphene nanopore. The results show that different nucleotides do indeed have distinguishable electronic signals in this idealized case. As explained above, in this simplified picture the relative order of conductance of the four nucleotides is dictated by their HOMO locations with respect to the graphene Fermi energy.

In the following we are going to investigate the impact that fluctuations will have on the measured conductance to see whether the above electronic signatures of different nucleotides could withstand such fluctuations. In principle, there are two major causes for fluctuation \cite{Lagerqvist2006}: (1) structure deformation of the highly flexible DNA polymer and (2) collision with water molecules and ions which are inevitable in the aqueous environment. Mathematically, the mentioned fluctuation could be accounted for by randomized $\vec{r}_i$ in \ref{tunneling_conductance} with a standard deviation $\sqrt{<\Delta\vec{r}_i^2>}$. Here the magnitude of $\sqrt{<\Delta\vec{r}_i^2>}$ characterizes the strength of fluctuation and in this work it was estimated by MD simulation (see Method and movies in the supporting material). $G(t)$ of overstretched ssDNA passing through the nanopore in the presence of fluctuation was then evaluated and plotted in \ref{fig3}(b). This figure illustrates that not only signatures of different bases, but also electrical characteristics of the gaps between bases remain discernible. The former, nominated as "base state", while the latter as "gap state" are marked in \ref{fig3}(b). Identification of gap state is a crucial requirement towards whole-strand sequencing \cite{Branton2008}. From a physics point of view, it requires that the change in conductance between two neighboring nucleotides is significant compared to experimental noise. Furthermore, the time scale for the gap state should be sufficiently large to be measurable. \ref{fig3}(b) shows that this could be achieved by incorporating over-stretch of the polymer into our BGLC system. Without the over-stretch, the electrical characteristics of the gap between bases would become blurred, making data separation of each base impossible (see Fig.S2 and accompanying discussion in supporting material).

Since the over-stretch of polynucleotides within the nanopore plays a fundamental role in our proposed sequencing strategy, we further explore the potential experimental implementation of it in our design. We suggest an electrical field approach in which the field required for DNA over-stretch within the pore is about $60$ pN/e $\approx 0.38$ V/nm \cite{Heng2005}. Without additional dragging force, under such a strong longitudinal electrical field the DNA translocation would be too fast for any practical measurement inside the pore \cite{Branton2008}. We propose that the nanopillars shown in \ref{fig1}(a) could be utilized to exert the required dragging force on the polymer to balance the electrical driving force: by tuning configuration and surface properties of the nanopillar array, the magnitude of the dragging force on the passing-by polymers could be manipulated within a large range \cite{Dorfman2010}. The main advantage of the above proposal is that the electrical field approach is a scaling-down approach for device integration compared to DNA stretching using optical or magnetic tweezers.

\begin{figure}
   \begin{center}
   \includegraphics*[width=12cm]{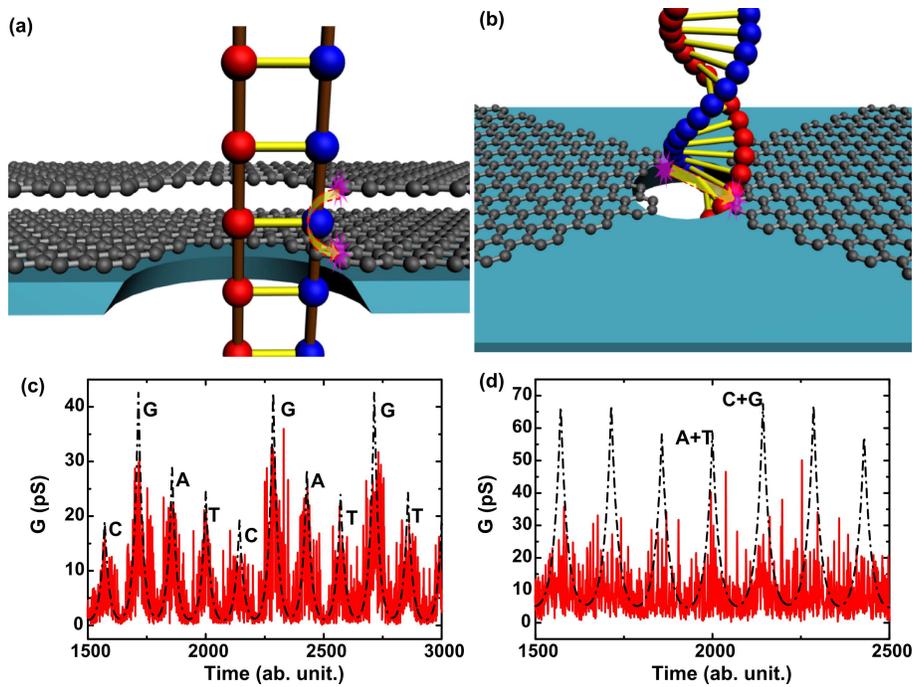}
   \caption{Sequencing double-stranded DNA with (a) bilayer graphene lateral contacts and (b) in-plane graphene electrodes. Coupling between the polymer and the contacts is characterized by pink lightening symbols. Directions of tunneling currents are schematically marked by yellow arrows. (c) and (d): Time-dependent tunneling conductance $G(t)$ of (a) and (b) respectively, where black dashed curves correspond to the ideal case, while red curves take fluctuations into account. While the four bases in one strand of the double-stranded DNA are identifiable in the bilayer graphene setup (c), the in-plane graphene electrode can as expected only distinguish between the AT base pair via its lower conductance level and the GC base pair via its higher conductance level (d).
   }
   \label{fig4}
   \end{center}
\end{figure}

Another potential application of our proposed device architecture is the possibility to sequence even dsDNA. If implemented, dsDNA sequencing would exhibit an overwhelming advantage over ssDNA sequencing in which the maximum strand length is severely limited by self-hybridization \cite{Tsutsui2010}. A careful inspection indicates that the obstacle of dsDNA sequencing with previous in-plane transverse electrodes lies in that each base-pair has to make contacts to both \emph{left} and \emph{right} probes to raise the transverse tunneling currents. This is clearly demonstrated in \ref{fig4}(b). Consequently, the electrical signal is the sum of contribution from two complementary nucleotides on that pair, resulting in an identification of merely the whole base pair (i.e., AT vs GC), but not capable of determining which base belongs to which strand. On the other hand, it may be feasible that in our system the nucleotides on only one strand are selected by the \textit{lateral} bilayer contacts during the translocation, as sketched in \ref{fig4}(a). By incorporating a certain molecule manipulation technology such as tethered to tweezers, one strand is maneuvered to be sufficiently close to the inner surface of the nanopore and hence gets detected by BLGC for electrical interrogation, while the other strand makes much more random contact and thus a trivial contribution to the overall conductance (a movie based on MD simulation is provided in the supporting material). This is quantitatively demonstrated in \ref{fig4}(c) and (d), where the calculated time-dependent tunneling conductance $G(t)$ of dsDNA in BLGC and 2-D transverse electrode systems are plotted respectively. The resolution may get further enhanced since our MD simulation reveals that upon strong manipulation, the force fluctuation $<\Delta \vec{r}^2>$ of the translocating strands gets remarkably attenuated \cite{Luan2011}. We emphasize that our computational results can only provide a tentative insight into the prospect of sequencing dsDNA with BLGC, although it seems very alluring.

\section{Conclusion}
In summary, we have proposed the use of adjacent layers of bilayer graphene embedded as two separate electrodes for detecting tunneling current when driving DNA polymer through a nanopore. Our theoretical study has shown that single-base resolution on the target DNA strand could be achieved readily. If the corresponding experiment is implemented successfully, nanopore sequencing with long-strand DNA could be performed while the complexity of the fabrication process is expected to be much more modest compared to other suggested device architectures.

\section{Method}

We performed MD simulation of single nucleotide dwelling in the bilayer graphene nanopore, single-stranded and double-stranded DNA translocating through the nanopore with NAMD2, and then extracted the time-variant atomic configuration of the target nucleotides. Details of MD simulation: the pore was made of two layers of graphene and with 2.4-nm thick silicon nitride material on the top and bottom as insulating layers; the pore diameter was about 4 nm; DNA molecules were constructed by NAMOT; The nanopore system was then solvated in TIP3 water with periodical boundary conditions in an NVT ensemble and with a 1 M solution of potassium and chlorine ions; the CHARMM27 force-field was used for DNA while UFF parameters were used for graphene carbon atoms and other atoms \cite{Lagerqvist2006, Krems2009}; a stretching force about 300 pN was exerted on the DNA strand and the electrical driving field $E_z \approx 1$ kcal/mol$\cdot$e$\AA$. The MD animation and associated time-variant nucleobase position $\vec{r}(t)$ were presented in Movie 1, 2 and 3 respectively. Based on these real-time information of molecular geometry, several important physical parameters were extracted: (1) Molecular orbitals of the under-scanning nucleotide was calculated by using YAEHMOP at each snapshot during the translocation; (2) the average HOMO positions $\varepsilon_{HOMO}$ of the four bases were then obtained; (3) $<\Delta\vec{r}^2>$ of the nucleotides during the translocation was calculated.

\acknowledgement
This research is supported partially by the Japan Society for the Promotion of Science (JSPS) through its "Funding Program for World-Leading Innovative R\&D on Science and Technology". RHS acknowledges financial support from the Swedish Foundation for International Cooperation in Research and Higher Education (STINT, Grant No. YR2010-7030) and the Swedish Research Council (VR, Grant No. 621-2009-3628).

\begin{suppinfo}
Movie 1 showing molecular dynamical simulation of single nucleotide dwelling in the bilayer graphene nanopore, and the associated fluctuation of molecule location. Movie 2 visualizing overstretched ssDNA threading through bilayer graphene nanopore and tracing position of the target nucleobase (No.4). Movie 3 demonstrating overstretched dsDNA passing by our nanopore and $\vec{r}(t)$ of complementary bases on the target pair (No.4). For clarity, water molecules, potassium and chloride ions and SiN membrane were not shown in the movies.
\end{suppinfo}

\bibliography{manuscript}

\clearpage

\end{document}